\documentclass{phb-proc4-auth}
\usepackage{graphicx}
\usepackage{bm}
\usepackage{xspace}
\usepackage{amssymb}
\newcommand{\urrs}{U(Ru$_{1-x}$Rh$_x$)$_2$Si$_2$\xspace }

\newcommand{\urs}{URu$_2$Si$_2$\xspace }
\newcommand{\tord}{$T_{\rm o}$\xspace}

\newcommand{\tMag}{$T_{\rm M}$\xspace}

\begin{document}
\begin{frontmatter}
\journal{SCES '04}
%
%
\title{
Magnetic Excitations in U(Ru$_{1-x}$Rh$_x$)$_2$Si$_2$ ($x\le 0.03$)
}

%
\author[Ibaraki]{M. Yokoyama,}
\author[Hokudai]{H. Amitsuka,}
\author[Hokudai]{S. Itoh,}
\author[Hokudai]{I. Kawasaki,}
\author[Hokudai]{K. Tenya,}
\author[ISSP]{H. Yoshizawa}

%
\address[Ibaraki]{Faculty of Science, Ibaraki University, Mito 310-8512, Japan}
\address[Hokudai]{Graduate School of Science, Hokkaido University, Sapporo 060-0810, Japan}
\address[ISSP]{Neutron Science Laboratory, Institute for Solid State Physics, The University of Tokyo, Tokai 319-1106, Japan}
%

\begin{abstract}
We have investigated magnetic excitations for a mixed phase of hidden order (HO) and the antiferromagnetic (AF) order in \urrs ($x\le 0.03$) by means of inelastic neutron scattering. The inelastic peaks observed at $Q=(1,0,0)$ and (1,0.4,0) in the HO phase for $x=0$ and 0.015 at 1.4 K are found to be strongly reduced in the AF dominant compositions of $x=0.02$ and 0.03. Similar behavior is observed as the HO is replaced by the AF order upon cooling for $x=0.02$. The $x-T$ region in which the strong reduction of inelastic peaks is observed corresponds to the AF-rich region, indicating that the magnetic excitations typical for the HO-phase vanish in the AF phase. 
\end{abstract}


%

\end{frontmatter}

%
The nature of the ordered state below $T_{\rm o}=17.5\ {\rm K}$ in \urs (the ThCr$_2$Si$_2$ type, body-centered tetragonal structure) \cite{rf:Palstra85} has been attracting renewed interest since the finding of the unusual evolution of the type-I antiferromagnetic (AF) phase under pressure $P$ \cite{rf:Ami99,rf:Ami2000}. Recent $^{29}$Si-NMR \cite{rf:Matsuda2001} and $\mu$SR \cite{rf:Ami2003} experiments under $P$ revealed that the evolution of the inhomogeneous AF phase is due to an effect of volume fraction, and indicates that the majority of the system is occupied by the ``hidden order" (HO) at ambient pressure below \tord . 

Quite recently, the elastic neutron scattering experiments performed on the Rh substitution system \urrs revealed that the $2\%$ substitution of Rh for Ru can also enhance the AF phase without applying $P$ \cite{rf:Yoko2004,rf:Bourdarot2003}. Except the suppression of both the phases at $x\sim 0.04$, the overall features are quite similar to that obtained from the pressure effect for the pure \urs . Because of no restriction of a pressure cell, the Rh dope system is expected to be suitable for the detailed microscopic investigation on the unusual two-phase competition. In this presentation, we report inelastic neutron scattering experiments on \urrs ($x\le 0.03$), for the first time, in order to study the low-energy magnetic excitations in this unusual mixed phase.

Single crystals of \urrs with $x=0,\ 0.015,\ 0.02$ and 0.03 were grown by the Czochralski method in a tetra-arc furnace, and vacuum-annealed at 1000$^{\circ}$C for 5 days. The samples with the volume of $\sim 150$ mm$^3$ were cut out of the ingots by means of the spark erosion, mounted in aluminum cans filled with $^4$He gas so that the scattering plane becomes $(hk0)$, and then cooled to 1.4 K in a $^4$He refrigerator. The inelastic neutron scattering measurements were performed on the triple-axis spectrometer GPTAS (4G) located at the JRR-3M reactor of the Japan Atomic Energy Research Institute. We made the constant-$Q$ scans at $Q=(1,0,0)$ and (1,0.4,0) with the fixed final momentum $k_{\rm f}=2.65\ {\rm \AA^{-1}}$. The combination of 40'-80'-40'-80' collimators and one pyrolytic graphite filter was chosen. The energy resolution determined from the FWHM of the vanadium incoherent scattering was $\sim$ 0.88 meV.

Figure 1 shows the magnetic-excitation spectra at 1.4 K obtained from the scans at $Q_{\rm a}=(1,0,0)$ and $Q_{\rm b}=(1,0.4,0)$. The contributions of instrumental background and incoherent scattering were carefully subtracted by scanning at the corresponding $|Q|$-invariant positions (0.707,0.707,0) and (0.762,0.762,0), where we observed neither magnon nor phonon scattering. The inelastic-scattering intensity at each $x$ is normalized by the integrated intensities of the nuclear Bragg scattering at (110). The sharp peak at the energy transfer $\hbar \omega \sim 0$ in the spectra at $Q_{\rm a}$ arises mainly from the higher-order nuclear Bragg reflections and the AF Bragg reflections. For pure \urs , clear peaks are observed at $\sim$ 2.4 meV ($Q_{\rm a}$) and at $\sim$ 4.6 meV ($Q_{\rm b}$), which are in good agreement with the previous results \cite{rf:Ami2000,rf:Broholm91}. By substituting 1.5\% of Rh for Ru, the position of the inelastic peak at $Q_{\rm a}$ shifts to $\sim$ 1.2 meV. Interestingly, the peak at $Q_{\rm a}$ suddenly disappears at $x=0.02$ and 0.03, where the AF phase with nearly full volume fraction replaces the HO phase \cite{rf:Yoko2004}. Such reduction is also seen in the $x$ variations of the spectra at $Q_{\rm b}$, but a heavily dumped peak is still observed at $x=0.02$ and 0.03 \cite{rf:Bourdarot2003}. 
\begin{figure}[tbp]
\begin{center}
\includegraphics[keepaspectratio,width=0.44\textwidth]{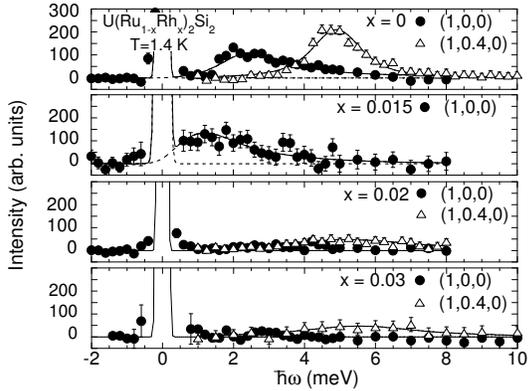}
\end{center}
\caption{The inelastic neutron scattering spectra at 1.4 K for \urrs ($x\le 0.03$), obtained from the constant-$Q$ scans at $Q=(1,0,0)$ and (1,0.4,0). Incoherent scattering and instrumental background are subtracted. The lines are guides to the eye.}
\end{figure}

The reduction of the inelastic peaks in the AF-rich phase is also observed in the temperature scans for $x=0.02$ (Fig.\ 2). We found a broad peak anomaly at $Q_{\rm a}$, with a peak position around 0.5 meV or lower, i.e.\ at clearly lower energy than in pure \urs . It appears below $\sim$ 13.5 K and grows significantly with decreasing temperature. Below 8 K, however, we found no significant anomaly in the whole energy range of $\hbar \omega \le 8\ {\rm meV}$. In our study using elastic neutron scattering and specific heat, \tord and the onset temperature \tMag of the large AF Bragg reflection are estimated to be $\sim$ 13.7 K and $\sim$ 8.3 K. The magnetic excitation at $Q_{\rm a}$ seen in the HO phase is thus considered to vanish in the AF phase. In contrast to the strong variations of the spectra at $Q_{\rm a}$, the magnitude of the heavily dumped peak observed at $Q_{\rm b}$ is insensitive to the temperature. It still exists above \tord with almost the same magnitude, suggesting that this magnetic fluctuation is not directly coupled to the two types of order.
\begin{figure}[tbp]
\begin{center}
\includegraphics[keepaspectratio,width=0.44\textwidth]{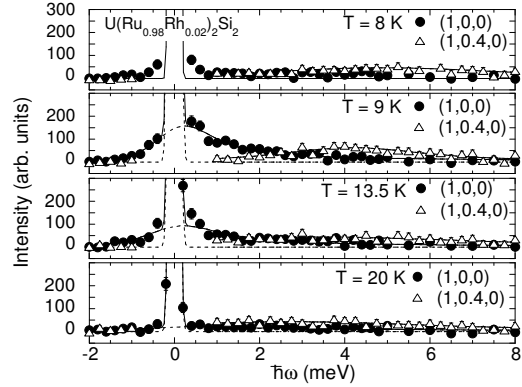}
\end{center}
\caption{Temperature variations of the constant-$Q$ scans at $Q=(1,0,0)$ and (1,0.4,0) in U(Ru$_{0.98}$Rh$_{0.02}$)$_2$Si$_2$. Incoherent scattering and background are subtracted. The lines are guides to the eye.}
\end{figure}

The disappearance of the magnetic excitations in the AF phase indicates that the matrix elements on the U 5f magnetic moment between the ground state and the low-energy excited states become zero in the AF order. This feature is consistent with the results for the pure \urs under $P$ \cite{rf:Ami2000,rf:Bourdarot2003}. A possible candidate for the HO parameters is the quadrupoles $\psi=J_x^2-J_y^2$ or $J_xJ_y+J_yJ_x$ originating in the non-Kramers doublet, since $\psi$ is orthogonal to the dipole moment $m=J_z$ bringing the AF order, and does not interact with the spin of neutrons \cite{rf:Ami2000,rf:Ohkawa99}.

In summary, the inelastic neutron scattering experiments on \urrs ($x\le 0.03$) showed that the magnetic-excitation peaks observed at $Q=(1,0,0)$ and (1,0.4,0) in the HO phase are strongly reduced in the AF-rich phase with increasing $x$ and decreasing $T$ ($x=0.02$), indicative of the disappearance of the spin-wave excitations in the AF phase. The further investigations with other momentum transfers are now in progress.

This work was supported by a Grant-in-Aid for Scientific Research from the Ministry of Education, Culture, Sports, Science and Technology of Japan.

\end{document}